\documentclass[%
 reprint,
 amsmath,amssymb,
 aps,
]{revtex4-2}

\usepackage{graphicx}
\usepackage{dcolumn}
\usepackage{bm}
\usepackage{xcolor}
\usepackage{hyperref}
\usepackage{CJKutf8}

\begin{document}
\begin{CJK}{UTF8}{gbsn} 
\preprint{APS/123-QED}

\title{Quantifying the non-Abelian property of Andreev bound states in inhomogeneous Majorana nanowires}

\author{Yu Zhang}
\affiliation{School of Physics, MOE Key Laboratory for Non-equilibrium Synthesis and Modulation of Condensed Matter, Xi’an Jiaotong University, Xi’an 710049, China}

\author{Yijia Wu}
\affiliation{Interdisciplinary Center for Theoretical Physics and Information Sciences, Fudan University, Shanghai 200433, China}
\affiliation{Hefei National Laboratory, Hefei 230088, China}

\author{Jie Liu}
\thanks{Corresponding author: jieliuphy@xjtu.edu.cn}
\affiliation{School of Physics, MOE Key Laboratory for Non-equilibrium Synthesis and Modulation of Condensed Matter, Xi’an Jiaotong University, Xi’an 710049, China}
\affiliation{Hefei National Laboratory, Hefei 230088, China}

\author{X. C. Xie}
\affiliation{Interdisciplinary Center for Theoretical Physics and Information Sciences, Fudan University, Shanghai 200433, China}
\affiliation{International Center for Quantum Materials, School of Physics, Peking University, Beijing 100871, China}
\affiliation{Hefei National Laboratory, Hefei 230088, China}

\date{\today}

\begin{abstract}
Non-Abelian braiding is a key property of Majorana zero modes (MZMs) that can be utilized for topological quantum computation. However, the presence of trivial Andreev bound states (ABSs) in topological superconductors can hinder the non-Abelian braiding of MZMs. We systematically investigate the braiding properties of ABSs induced by various inhomogeneous potentials in nanowires and quantify the main obstacles to non-Abelian braiding. We find that if a trivial ABSs is present at zero energy with a tiny energy fluctuation,
 their non-Abelian braiding property can be sustained  for a longer braiding time cost, since the undesired dynamic phase is suppressed. Under certain conditions, the non-Abelian braiding of ABSs can even surpass that of MZMs in realistic systems, suggesting that ABSs might also be suitable for topological quantum computation.

\end{abstract}

\maketitle


\section{\label{sec:level1}Introduction}

The semiconductor-superconductor nanowire has been deemed as one of the most promising platforms for pursuing Majorana zero modes (MZMs) \cite{sau, fujimoto, sato, alicea2, lut}. 
 After the first semiconductor-superconductor nanowire system being fabricated in 2012\cite{kou}, such systems have been fabricated in a number of experiments \cite{deng, das1, hao1, Marcus, deng2, perge, Yaz2, PJJ1, PJJ2, QZB1}. However,  none of them can convincingly demonstrate the existence of MZMs in such systems.
The central issue is that trivial Andreev bound states (ABSs) can 
 mimic the signal of MZMs \cite{Jie1, brouwer, Aguado,AguadoN,ABSM,Moore,ChunXiao1,Aguado2,Wimmer,Tewari, 075161, 184520, 035312, 155314,075416, 124001, 631031, 104273}, making them difficult to be distinguished using standard conductance measurement methods.
These ABSs are induced by inhomogeneous potential or disorder which are difficult to eliminate with the state of art nanotechnology.  
Unlike the MZMs, which are nonlocally distributed at both ends of the nanowire, these states are spaticaly partially-separated in space and exhibit less topological protection. 
In principle, non-local conductance measurements could be a powerful method to distinguish these two types of states\cite{ABS1, ABS2, ABS3,ABSM1, ABS4, ABS5, ABS6, ABS7,ABS8, ABS9, Klinovaja,MR1,MR2} . 
However, other possibilities are still hard to be ruled out \cite{NL1,NL2,NL3}.

The most fundamental way to distinguish MZM from the ABSs is certainly based on its non-Abelian statistics \cite{Ivanov, alicea3, NQP,Jienew,TQC2, TQC3, TQC4}.  
When two MZMs are spatially exchanged, they obey an unusual rule  $\gamma_i\rightarrow\gamma_j$, and $\gamma_j\rightarrow-\gamma_i$. This braiding would rotate the degenerate ground state
formed by MZMs and constitute the basic logic gate of the quantum computation. 
Thus, investigating the braiding properties of MZMs in the presence of ABSs not only can distinguish the MZMs from the ABSs in principle, but also is a necessary way towards the topological quantum computation.
Although several groups have suggested that ABSs may also be utilized for non-Abelian braiding\cite{AguadoN,Wimmer}, direct investigations on their non-Abelian braiding properties are still lacking. Previously, we revealed that ABSs with finite energy could introduce an additional dynamic phase that disrupts non-Abelian braiding\cite{JieA}. The key factors detrimental to this process remain an open question.
 Furthermore, ABS can be induced by various effects, such as inhomogeneities in chemical potential, superconducting strength, or disorder\cite{Tewari, 075161, 054510, 013377, 124001}.
 How to quantify the difference of these ABSs in different mechanism is certainly necessary.
 This not only provides a perspective for distinguishing MZMs from ABSs from the viewpoint of non-Abelian braiding but also sheds light on how to recover non-Abelian braiding in the presence of ABS.
 
\begin{figure}
\includegraphics[width=3in]{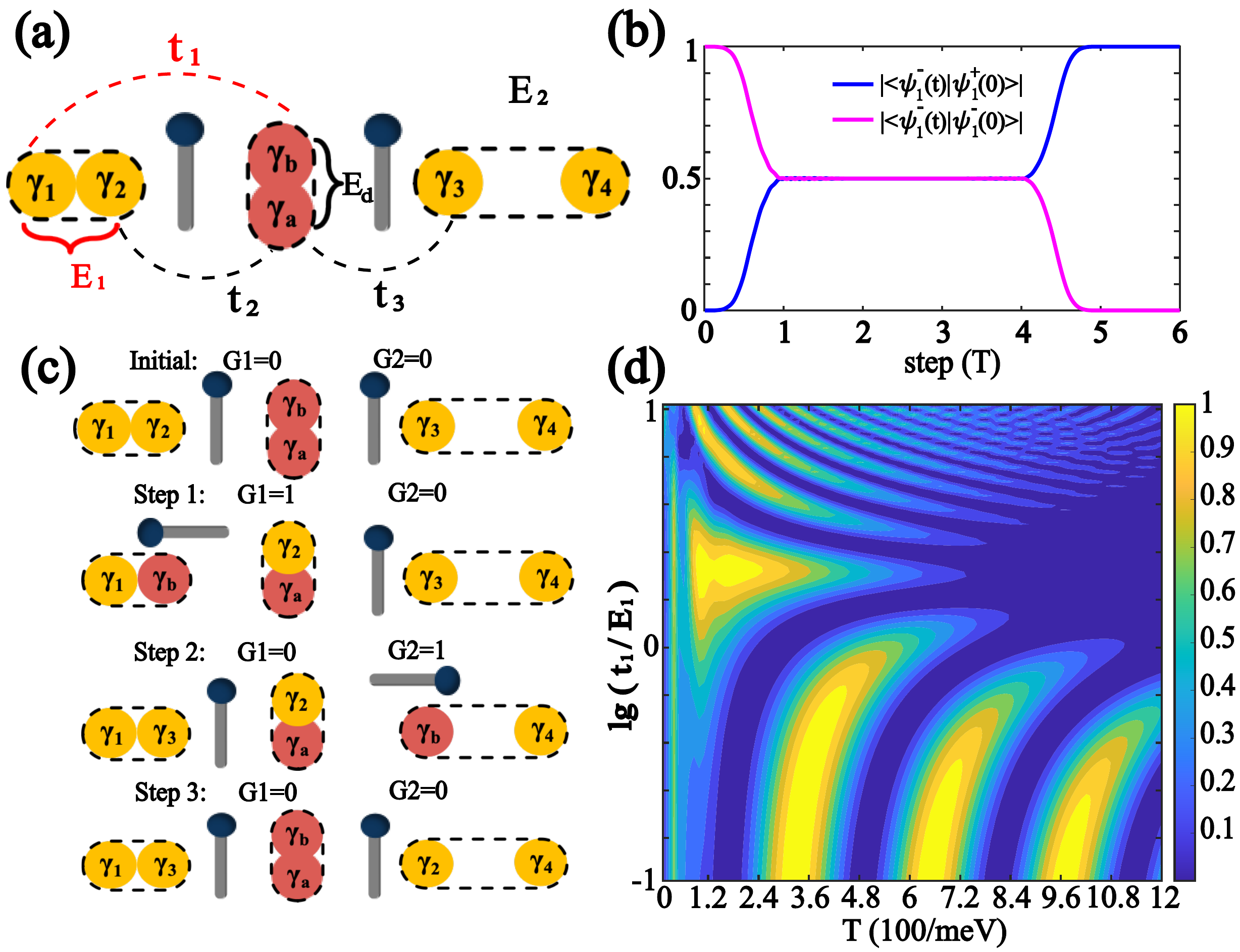}
\caption{\label{fig1} 
(a) The schematic  plot of a minimal effective model for braiding.  MZMs can be moved and swapped with the assistance of the QD  by manipulating the coupling strengths.
(b) Evolution of the wave function $\psi_1 ^{-}(t)$ for a perfect MZMs with $E_1=0$ and $t_1=0$. It evolves into $\psi_1 ^{+}(0)$ after swapping $\gamma_2$ and $\gamma_3$  twice in succession. Here we set $t_2 = t_3 = 0.3meV$, $E_d = 0.3meV$.
(c) Braiding operation steps for non-abelian braiding. We can exchange $\gamma_2$ and $\gamma_3$ through turning on and off of $G_1$ and $G_2$.
(d) The braiding results as a function of $T$ with the influence of $t_1$ and $E_1$. Here we  fix $E_1 = 0.01meV$. The other parameters are the same as (b) .}
\end{figure}

 In this manuscript, we try to quantify the braiding properties of these ABSs in various situations.  
We primarily focus on nearly zero-energy ABSs, which display quantized zero-bias peaks during conductance measurements, 
making them  difficult to be distinguished from MZMs through conductance measurements.
 We find that these nearly zero-energy ABSs can be deemed as two weakly coupled MZMs in a short nanowire.
 As shown in Fig. \ref{fig1}(a),  a nearby MZM $\gamma_1$ can hybridize with a  MZM $\gamma_2$ being exchanged, inducing a weak coupling $E_1$.
  Moreover, it can also induce an additional coupling $t_1$ with the auxiliary quantum dot (which is composed of $\gamma_a$ and $\gamma_b$).
 Both $E_1$ and $t_1$  could induce a dynamic phase and ruin the non-abelian braiding. Interestingly,  $t_1$ is closely related with $E_1$ in these nearly-zero-energy ABSs during the braiding. They oscillate with almost the same order of  oscillation period.
 That is to say, if  $E_1$ remains stable around zero energy with tiny energy fluctuations, $t_1$ will also be very small in that region. Then the braiding properties of these ABSs are similar to those of true MZMs.
  This suggests that nearly zero-energy ABSs could potentially be used for topological quantum computation. In fact, we find that under certain conditions, these ABSs might even outperform the true MZMs during the braiding. 
  This may shield some light on topological quantum computation in the presence of ABS.

The rest of this article is organized as follows. In section \Ref{sec:level2} , we introduce the Hamiltonian of a low-energy effective braiding model, and show the results of braiding with the influence of $E_1$ and $t_1$. 
In Section  \Ref{sec:level3}, we introduce the tight binding model of a real semiconductor superconductor nanowire.  Then we investigate the braiding property of ABSs in various situations. In subsection \Ref {sec:A} we investigate the braiding property of MZMs with the influence of length effect.  In subsection \Ref {sec:B} we show the braiding property of ABSs induced by inhomogeneous chemical potential. In  subsection \Ref {sec:C} we investigate braiding property of ABS caused by quantum-dot like structure, and in subsection \Ref {sec:D}  we
show the influence of disorder.  While in  subsection \Ref {sec:E} we show that the non-abelian braiding may be modified by inhomogeneous potential.
Finally in section  \Ref{sec:5}  we give a conclusion on the braiding property of ABSs.

\section{\label{sec:level2} Effective model for braiding ABSs}
To investigate the braiding properties in the presence of ABSs, we follow the previous proposal in which the MZMs are braided with the assistance of a quantum dot (QD) \cite{Jienew}. 
Such a  structure is actually constructed by adding an additional QD  to the Josephson junction.  As shown in Fig. \ref{fig1}(a),  the right side is an ideal semiconductor superconductor nanowire with a pair of MZMs  non-locally distributed at the two ends of the nanowire. To facilitate the analysis, we assume these two MZMs $\gamma_3$ and $\gamma_4$ are well separated,  the coupling energy $E_2$ between them is almost zero. In contrast, the left side is an inhomogeneous or disordered nanowire with the  ABSs stays at the end. 
Our simulations in concrete model suggest that  these nearly zero energy ABSs  can be deemed as a pair of finite-overlapped MZMs $\gamma_1$ and $\gamma_2$,  with the coupling energy $E_1$.  
What is more, $\gamma_2$ and $\gamma_3$  would sequentially  couple to the assisted QD by turning off the correspondingly  gate as revealed by Fig. \ref{fig1}(c).  $\gamma_1$ will also couple to the QD when we turn off the gate $G_1$. 
Thus, the whole effective Hamiltonian describing such setup is given by
\begin{equation}
\begin{split}
H_E(t)=2 E_d d^{\dagger} d+\mathrm{i} E_1 \gamma_1 \gamma_2+\mathrm{i}\left[t_2(t) d+t_2(t)^* d^{\dagger}\right] \gamma_2\\
+\left[t_1(t)^* d^{\dagger}-t_1(t) d\right] \gamma_1+\left[t_3(t) d-t_3(t)^* d^{\dagger}\right] \gamma_3
\end{split}
\end{equation}
Here, $d$ is the annihilation operator for the fermionic state in the QD, and $E_d$ is the on-site energy of this QD state.  In addition, the coupling strength between QD and $\gamma_1$,  $\gamma_2$, and  $\gamma_3$ are $t_1(t)=\left|t_1(t)\right| \mathrm{e}^{\mathrm{i} \phi_1 / 2}$,  $t_2(t)=\left|t_2(t)\right| \mathrm{e}^{\mathrm{i} \phi_1 / 2}$ and $t_3(t)=\left|t_3(t)\right| \mathrm{e}^{\mathrm{i} \phi_2 / 2}$, respectively. 
Here the phases $\phi_1$ and $\phi_2$ are the superconducting phases of the two nanowires respectively. 
Following the previous proposal, we set $\phi_1-\phi_2=\pi$ to facilitate the braiding. What is more, we can express the QD state in terms of the two Majorana operators $\gamma_a$ and $\gamma_b$, as $d=\frac{1}{2} \mathrm{e}^{-\mathrm{i} \frac{\phi_1}{2}}\left(\gamma_a+\mathrm{i} \gamma_b\right)$. Therefore, the above Hamiltonian can be further expressed by the Majorana operators as:

\begin{equation} \label{Eq2} 
\begin{aligned} 
H_{EM}(t)= & \mathrm{i} E_d \gamma_a \gamma_b+\mathrm{i} E_1 \gamma_1 \gamma_2+\mathrm{i}\left|t_2(t)\right| \gamma_a \gamma_2-\mathrm{i}\left|t_1(t)\right| \gamma_b \gamma_1 \\ 
& -\mathrm{i}\left|t_3(t)\right|\gamma_a \gamma_3
\end{aligned}
\end{equation}

 \begin{table*}
\caption{\label{Tab1}  Parameters for potential and superconducting pairing amplitude in different physical mechanism.}
\begin{ruledtabular}
\begin{tabular}{cccccc}
 physical mechanism&$\mu$(meV)&V(x)(meV)&$\Delta$(meV)&$\alpha$(meV $\cdot$ nm)& L($\mu m$)\\ \hline
 Uniform nanowire&0&0 &0.25&40&1.5 \\
 
 Step-like IP&1.2&$V(x) = \begin{cases}0 & \text { if } x<2.3 \\ 1.2 & \text { if } x \geq 2.3\end{cases}$&0.25&40&2.5\\
 
 Smooth IP&1.5&$V(x) = \begin{cases}0 & \text { if } x<1.875 \\ 1.5\sin(\frac{(x-1.875)\pi}{1.25}) & \text { if } x \geq 1.875\end{cases}$& 0.25 &40&2.5\\
 
 Step-like QD &0.4& $V(x) = \begin{cases}0 & \text { if } x<2.3 \\ 0.4 & \text { if } x \geq 2.3\end{cases}$
 &$\Delta(x) =\begin{cases}0.25 & \text { if } x<2.3 \\ 0 & \text { if } x \geq 2.3\end{cases}$
 &40&2.5\\
 
 Smooth QD &1&$V(x) = \begin{cases}0 & \text { if } x<1.75 \\ 1\sin(\frac{(x-1.75)\pi}{1.5}) & \text { if } x \geq 1.75\end{cases}$
 &$\Delta(x) =\begin{cases}0.25 & \text { if } x<1.75 \\ 0 & \text { if } x \geq 1.75\end{cases}$
 &40&2.5\\
 
 Disorder&1& $V(x) \sim \mathcal{N}\left(0,1^2\right)$ &0.25& 40& 2.5\end{tabular}
\end{ruledtabular}
\end{table*}

If we set $E_1$ and $t_1(t)$ to zero, then this Hamiltonian can return to the previous Y-junction minimal model for braiding MZMs. The operation  is also the same as before\cite{Jienew,Li}, which can be concluded as three steps (the time-cost for each step is  $T$ ). Initially,  the junction is disconnected with each other, that is, $t_2(0) = t_3 (0)= 0$ and $E_d$ is set to a non-zero initial value $E_0$.  In step 1, we turn off the gate $G_1$, then $t_2$ is increased  as  $t_2(t) = \frac{1-\cos \left(\frac{t}{T} \pi\right)}{2} t_c$ and simultaneously $E_d$ is turned to zero as $E_d(t)=\frac{1+\cos (\frac{t}{T} \pi)}{2} E_0$ through gate voltage  ( not shown in the figure). In step 2, $|t_3(t)|$ is increased from $0$ to $t_c$ as $t_3(t) = \frac{1-\cos \left(\frac{t}{T} \pi\right)}{2} t_c$ by turning off the gate  $G_2$, simultaneously we turn on the gate  $G_1$, then  $t_2$ is decreased 
as $t_2(t) = \frac{1+\cos \left(\frac{t}{T} \pi\right)}{2} t_c$. 
Finally in step 3, we turn on the gate $G_2$, $|t_3|$ is decreased to $0$ as  $t_3(t) = \frac{1+\cos \left(\frac{t}{T} \pi\right)}{2} t_c$, and $E_d$ is returns back to initial value as $E_d(t)=\frac{1-\cos (\frac{t}{T} \pi)}{2} E_0$. 
Then all the parameters come back to their initial forms, while $\gamma_2$ is  exchanged with $\gamma_3$ with the exchange rule $\gamma_2\rightarrow\gamma_3$, $\gamma_3\rightarrow-\gamma_2$. 
For two pairs of MZMs whose eigenstates are in the wavefunctions of $\psi_j^{\pm}(0) = (\gamma_{2j-1}\pm i\gamma_{2j})/\sqrt{2}$ ($j = 1,2$).
If $\gamma_2$ and $\gamma_3$ are swapped twice in succession, then the wavefunction will evolve into $\psi_1^{\pm} (6T) =  (\gamma_1\mp i\gamma_2)/\sqrt{2}=\psi_1^{\mp}(0)$ and $\psi_2^{\pm} (6T) =  (-\gamma_3\pm i\gamma_4)/\sqrt{2}=-\psi_2^{\mp}(0)$
as shown in Fig. \Ref{fig1}(b).  Such evolution result demonstrates the non-abelian braiding properties of MZMs.

As a contrast for the ABSs state, such results  no longer hold true since $E_1$ and $t_1$ are not strictly zero. 
Following the results of  Ref. [\onlinecite{JieA}],  the wavefunction will evolute as $\psi(t) = U(t)\psi(0)$ with $U(t) = \hat{T}\exp [-i\int dt (E_{1,eff}(t)\gamma_1\gamma_2+t_{1,eff}\gamma_3\gamma_1+\Omega\gamma_2\gamma_3)]$.  
Here $\Omega$   is the geometric phase which is independent of the time cost $T$.  It is exactly $\pi/2$ if we exchange $\gamma_2$ and $\gamma_3$ twice.
In an even or odd parity qubit space, we can define the qubit operator\cite{TQC4} as $\sigma_x = -i\gamma_2\gamma_3$, $\sigma_y = -i\gamma_3\gamma_1$, $\sigma_z = -i\gamma_1\gamma_2$.
 Then  $ \exp (-i\int dt\Omega\gamma_2\gamma_3)= \exp (\frac{\pi}{2}\sigma_x)$ is exactly a NOT gate which rotates $\psi_j^{\pm}$ to $\psi_j^{\mp}$.
 Meanwhile,  $E_{1,eff}$  and $t_{1,eff}$ are the effective dynamic phase induced by the additional coupling terms $E_1$ and $t_1$, correspondingly. They will accumulate during the braiding  and become dominant when the braiding time cost $T$ is large.  
 Figure \ref{fig1} (d) numerically calculate the weight of $\psi_1^{-} (6T)$ on $\psi_1^{+}(0)$ versus the time cost $T$ and $t_1$, we fix $E_1$ as 0.001meV during the braiding. Since $t_1$ is  manipulated by the gate $G_1$. It varies as 
$t_1(t) =  \frac{1-\cos \left(\frac{t}{T} \pi\right)}{2} t_{1}$ when we turn off the gate and $t_1(t) =  \frac{1+\cos \left(\frac{t}{T} \pi\right)}{2} t_{1}$ when we turn on the gate $G_1$. 
Clearly, $|\langle\psi_1^{-} (6T)|\psi_1^{+}(0)\rangle|$ would approach to $1$ despite the influence of non-adiabatic effect if $T$ approaches zero, since $\Omega$ term dominates  in this situation. 
As $T$ increases, the dynamic phase becomes dominant.  If $t_1$ is small, the $E_1$ term dominates and rotates the qubit space along the $z$ axis,  causing $|\angle\psi_1^{-} (6T)|\psi_1^{+}(0)\rangle|$ to oscillate over time. 
As $t_1$  increases, the oscillation period also increases, but the amplitude quickly dampens to zero due to the non-commuting nature of $E_{1,eff} \sigma_z$ and $t_{1,eff} \sigma_y$.  
As a contrast, when $t_1\gg E_1$,  the $t_1$ term is dominant and rotates the qubit space along the $y$-axis, leading to rapid oscillations of the weight due to the large $t_1$.
In summary, the weight oscillates with time when only $E_1$ or $t_1$ is present, but it dampens due to their non-commuting nature.
 Finally, we want to emphasize that the dynamical behavior induced by $E_1$ and $t_1$ may vary for different paths.   
However, when the path is fixed, the dynamical behavior is primarily related to the ratio of 
$E_1$ and $t_1$. Thus we can quantify $E_1$ and $t_1$ by analyzing the oscillatory behavior of the wave function.

\begin{figure*}
\includegraphics[width=7in]{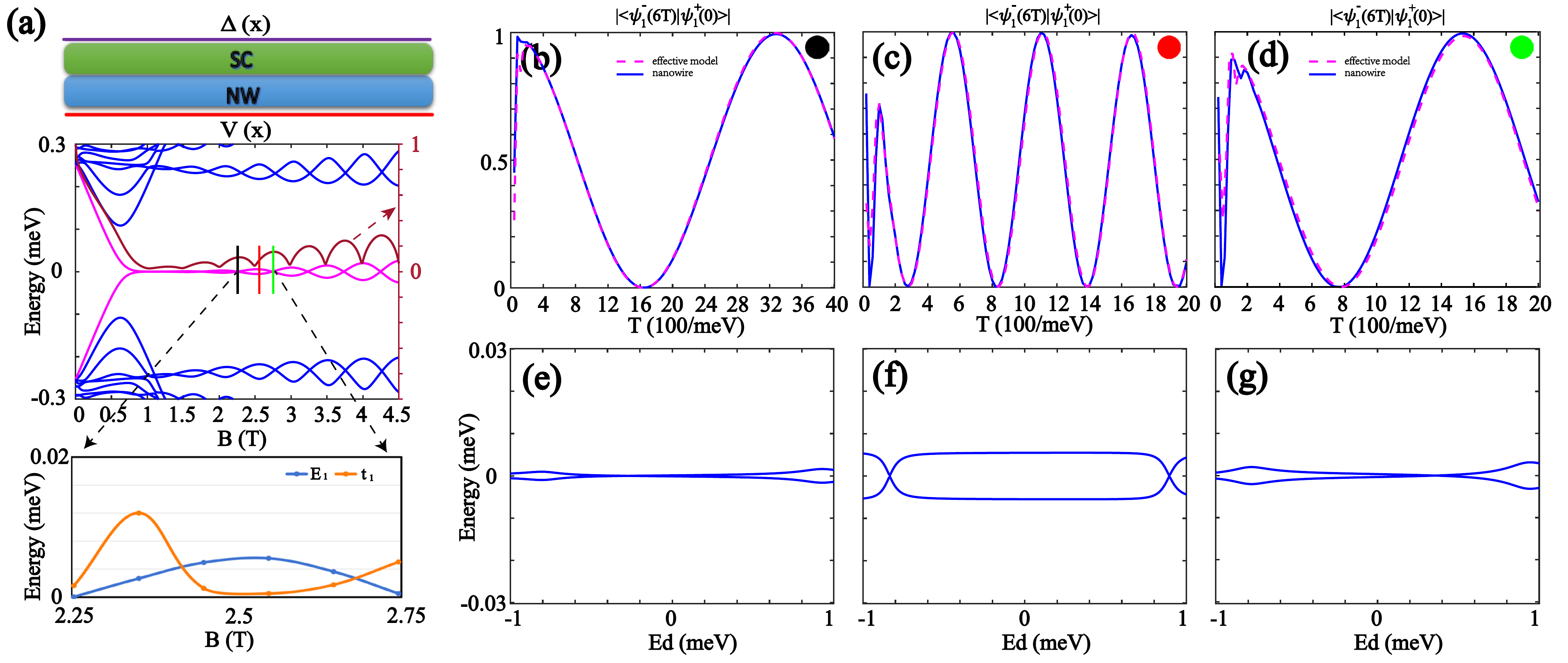}
\caption{\label{fig2}
Uniform nanowire with finite length. (a) Top panel shows the schematic plot of the uniform nanowire model. Here, blue solid line  is the distribution of $\Delta(x)$ along the wire, while the red solid line is the correspond inhomogeneous chemical potential along the wire which described by the first row of Table \ref{Tab1}.
 Middle panel shows the corresponding energy spectrum, and the dark red line shows the corresponding local estimator $\eta$. Bottom panel shows the value of $E_1$ and the corresponding effective $t_1$ which focus on the range from $B=2.25T$ to $B=2.75T$. 
(b)-(d) show the typical braiding results as a function of braiding time cost $T$，which corresponds to the values of $B_x$ marked by the vertical lines of different colors in the energy spectrum of (a)(indicated by the solid line) and the dashed line presents the fitting result of effective model.  
 (e)-(g) show the energy spectrum of quantum dot-nanowire model as the function of the on-site energy of QD state $E_d$ with different $B_x$, which corresponds to the values of $B_x$ marked by the vertical lines of different colors in the energy spectrum of (a).
 In the (b) and (e), $B_x = 2.25 T$, $E_1 = 3.368\times 10^{-5} meV$ and the corresponding effective $t_1 = 2.00\times 10^{-3}meV$. 
 In the (c) and (f), $B_x = 2.55 T$, $E_1 = 5.500\times 10^{-3} meV$ and the corresponding effective $t_1 = 2.00\times 10^{-4}meV$.
 In the (d) and (g), $B_x = 2.75 T$, $E_1 = 4.885\times 10^{-4} meV$ and the corresponding effective $t_1 = 5.10\times 10^{-3}meV$.}
\end{figure*}

\section{\label{sec:level3} numerical simulation in real  semiconductor superconductor nanowire}
Since ABSs are caused by inhomogeneous potential or disorder in semiconductor superconductor nanowire,  we need to introduce a realistic system to simulate the ABS.
The  Hamiltonian for the whole  system is given as follows:
\begin{equation}
\begin{aligned} 
H_{T}(t)=  \sum\nolimits_{i=1,2}H_{Si}+H_C(t)+H_{D}
\end{aligned}
\end{equation}

Here $H_{S1(2)}$ is the tight binding Hamiltonian of the semiconductor superconductor nanowire which stays at the left (right) side of the QD, it is given by
\begin{eqnarray}
\label{model}
 H_{Si} &=& \sum\nolimits_{\mathbf{x},\alpha} { - t_0(\psi _{i\mathbf{x} + \mathbf{a},\alpha }^\dag \psi_{i\mathbf{x},\alpha } + h.c.) - \mu\psi_{i\mathbf{x},\alpha }^\dag \psi_{i\mathbf{x},\alpha } } \nonumber \\
&+& \sum\nolimits_{\mathbf{x}, \alpha ,\beta } { - i{U _R} \psi _{i\mathbf{x} + \mathbf{a},\alpha }^\dag  \hat z \cdot (\vec{\sigma_y} )^{\alpha \beta }   \psi _{i\mathbf{x},\beta } } \nonumber \\
&+& \sum\nolimits_{\mathbf{x},\alpha} \Delta(x)e^{i\phi} \psi _{i\mathbf{x},\alpha }^{\dagger} \psi _{i\mathbf{x},-\alpha }^{\dagger} +V(x)\psi_{i\mathbf{x},\alpha }^\dag \psi_{i\mathbf{x},\alpha }  \nonumber \\
&+& \sum\nolimits_{\mathbf{x},\alpha ,\beta } { \psi _{i\mathbf{x}, \alpha }^\dag (g\mu_BB_x \vec{\sigma}_x)_{\alpha \beta} \psi _{i\mathbf{x}, \beta}}+ h.c..
\end{eqnarray}

\noindent Here, the subscript $\mathbf{x}$ denotes the lattice site; $\mathbf{a}$ is the lattice constant; $\alpha$ and $\beta$ are the spin indices;
 $t_0=h^2/(2m^{*}a^2)$ denotes the hopping amplitude;  $\mu $ is the chemical potential of the wire; 
$U_{R}=\alpha/(2a)$ is the Rashba coupling strength; and $g\mu_BB_x$ is the Zeeman energy,  with $g$ the effective $g$ factor, $\mu_B$ the Bohr magneton and $B_x$ the magnetic field.  
 $\Delta(x)$ is the superconducting pairing amplitude in site $x$.  Here we keep $\Delta(x)$ as a constant value $\Delta$ in the right side of the nanowire, and it may vary for different sites in the left side of the nanowire since it is an inhomogeneous wire.  Based on the same reason, 
 we introduce $V(x)$ to describe the inhomogeneous chemical potential or disorder in site $x$.  It is zero in the right side of the nanowire, while varies for different sites in the left side of the nanowire. 
  As shown in the table \ref{Tab1} ,  we give the parameters for five typical  types of inhomogeneous case which can induce nearly zero energy  ABSs.
 The other parameters are given as follows: the effective mass  $m^{*}=0.026m_e$, the Rashba spin-orbit coupling strength $\alpha=30\text{ meV}\cdot \text{nm}$, $\Delta=0.25 \text{ meV}$, $g=15$,  the lattice constant $a = 25 \text{nm}$.

 $H_d$ is the Hamiltonian of the QD, we introduce a Zeeman term to break the dengency of the spin.
\begin{equation}
 	 H_d=\sum\nolimits_{\alpha}2E_dd^\dagger_\alpha d_\alpha+\sum\nolimits_{\alpha ,\beta } { d _{ \alpha }^\dag (V_x\vec{\sigma}_x)_{\alpha \beta} d _{ \beta}},
\end{equation}

$H_{C}(t)$ describe the coupling between the end of two nanowires and the QD .
\begin{equation}
	H_{C}(t)=\sum\nolimits_{i=1,2;\alpha}C_{i}(t)\psi^\dagger_{i,x_{end},\alpha}d_{\alpha}+h.c.,
\end{equation}
\noindent here $C_{1}(t)$ and $C_{2}(t)$ is the coupling strength which can be tuned by  gate $G_1$ and $G_2$. 
Following the same proposal introduced in the section of effective model,  $C_{1(2)}(t)= \frac{1-\cos \left(\frac{t}{T} \pi\right)}{2} t_c$ when we turn off gate $G_1$($G_2$) and $C_{1(2)}(t)= \frac{1+\cos \left(\frac{t}{T} \pi\right)}{2} t_c$ when we turn on gate $G_1$($G_2$). In the numerical simulation of the realistic system, $t_c$ is set to a fixed value with $t_c = 0.23t_0$. 
We should notice there are three parameters $t_1(t)$, $t_2(t)$ and $t_3(t)$ which are controlled by gates in effective model of Eq. \ref{Eq2}.  While only two parameters are needed in real system, since $t_1(t)$ and $t_2(t)$ are controlled by the same gate $G_1$.
\begin{figure*}
\includegraphics[width=7in]{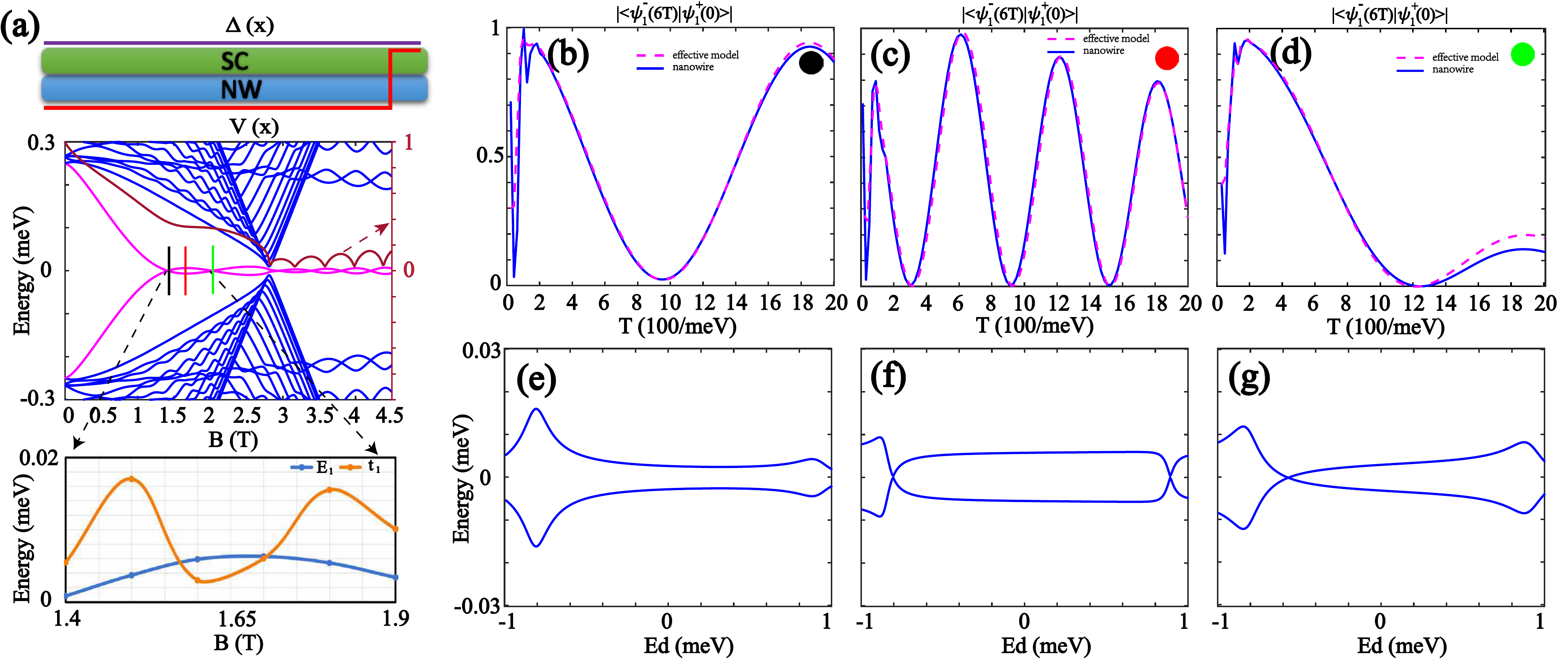}
\caption{\label{fig3}
Step-like inhomogeneous potential. (a)Top panel show the schematic plot of the step-like inhomogeneous potential model . The middle panel shows the energy spectrum, and the dark red line shows the local estimator $\eta$. The bottom panel shows that the value of $E_1$ and $t_1$ oscillate as the functions of $B_x$. 
(b)-(d) show the braiding results as a function of braiding time  cost at the different $B_x$，which corresponds to the values of $B_x$ marked by the vertical lines of different colors in the energy spectrum of (a)(indicated by the solid line) and the dashed line presents the fitting result by effective model.  
 (e)-(g) show the energy spectrum of QD-nanowire model as the function of the on-site energy of QD state $E_d$ at the different $B_x$, which corresponds to the values of $B_x$ marked by the vertical lines of different colors in the energy spectrum of (a).
 In the (b) and (e), $B_x = 1.4 T$, $E_1 = 8.289\times 10^{-4} meV$ and $t_1 = 5.10\times 10^{-3}meV$. 
 In the (c) and (f), $B_x = 1.6 T$, $E_1 = 5.900\times 10^{-3} meV$ and $t_1 = 2.00\times 10^{-3}meV$.
 In the (d) and (g), $B_x = 1.9 T$, $E_1 = 3.400\times 10^{-3} meV$ and $t_1 = 9.25\times 10^{-3}meV$.
 The other parameters are shown in the second row of Table 1.
 }
\end{figure*}

\begin{figure*}
\includegraphics[width=7in]{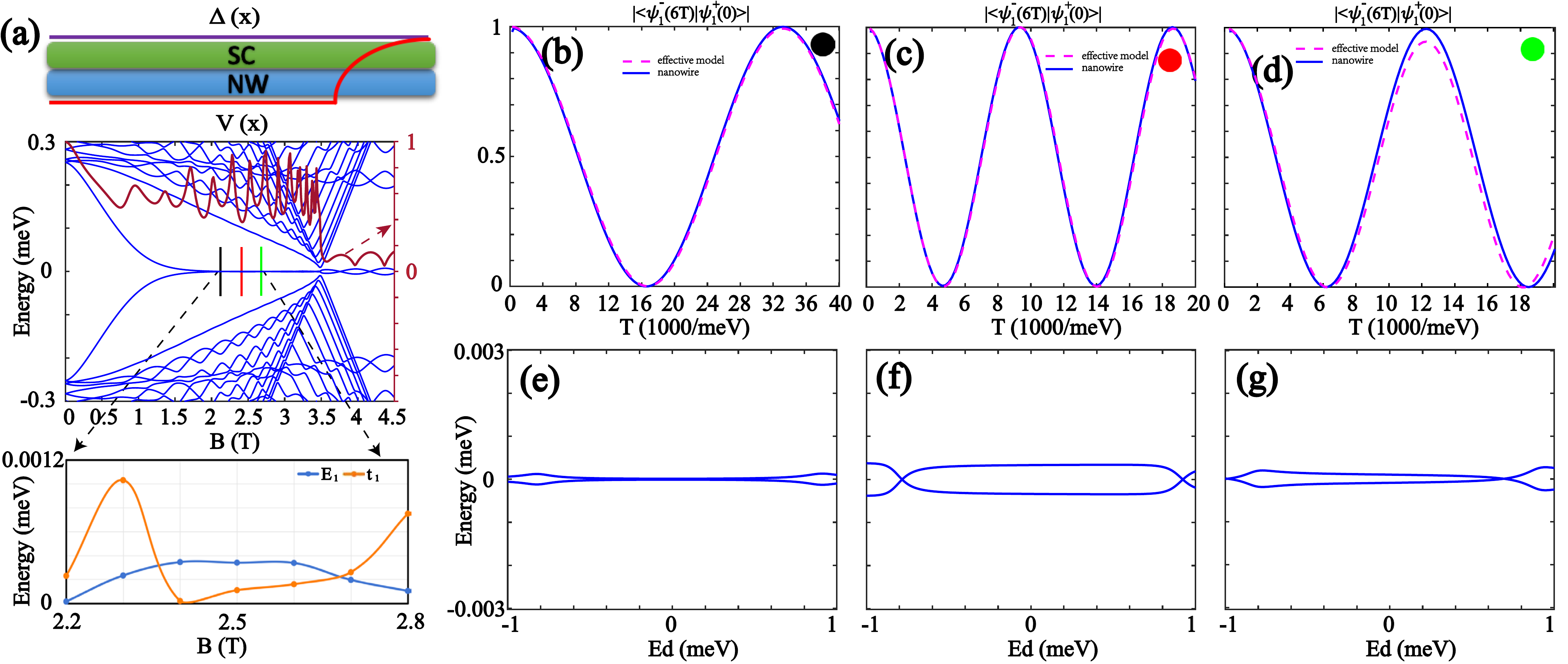}
\caption{\label{fig4}
Smooth inhomogeneous potential. (a) Top panel shows the schematic plot of the smooth inhomogeneous potential model. The middle panel shows the energy spectrum, and the dark red line shows the local estimator $\eta$. 
The bottom panel shows that the value of $E_1$ and $t_1$ oscillating as the functions of $B_x$. 
(b)-(d) show the braiding results as a function of braiding time cost at the different $B_x$，which corresponds to the values of $B_x$ marked by the vertical lines of different colors in the energy spectrum of (a)(indicated by the solid line) and the dashed line presents the fitting result by effective model.  
 (e)-(g) show the energy spectrum of QD-nanowire model as the function of the on-site energy of QD state $E_d$ at the different $B_x$, which corresponds to the values of $B_x$ marked by the vertical lines of different colors in the energy spectrum of (a).
 In the (b) and (e), $B_x = 0.6900 T$, $E_1 = 1.3569\times 10^{-4} meV$ and $t_1 = 4.60\times 10^{-3}meV$. 
 In the (c) and (f), $B_x = 0.8775 T$, $E_1 = 2.1000\times 10^{-3} meV$ and $t_1 = 9.00\times 10^{-4}meV$.
 In the (d) and (g), $B_x = 1.0650 T$, $E_1 = 3.2964\times 10^{-4} meV$ and $t_1 = 1.51\times 10^{-3}meV$.
 The other parameters are shown in the third row of Table \ref{Tab1}.
 }
\end{figure*}

\subsection{\label{sec:A} Uniform nanowire with finite length}
Here, we reveal how the braiding properties of MZMs are influenced by the length of the nanowire. 
In a uniform nanowire, the coupling energy of MZMs decreases exponentially with the increasing of the length of the nanowire.
 If the nanowire is not long enough, the two MZMs localized at the ends will inevitably hybridize.
  In the middle panel of Fig. \ref{fig2}(a), we plot the energy spectrum versus the Zeeman field for a short nanowire, with its parameters indicated in the first row of Table \ref{Tab1}.
 The energy spectrum of MZMs is nearly zero energy  when the system enters into the topological region due to the finite length effect.  
 However, if we focus on a typical region range from $B=2.25T$ to $B=2.75T$, the energy spectrum actually oscillates around zero energy, as indicated by the blue line in the bottom panel of Fig. \ref{fig2}(a).
Following the proposal outlined in Section II, we investigate the braiding properties in this specific region.
Fig. \ref{fig2}(b)-(d) show the final weight of  $|\langle\psi_1^{-} (6T)|\psi_1^{+}(0)\rangle|$ versus the  braiding time costing $T$ at $B_x=2.25T$, $B_x=2.55T$ and $B_x=2.75T$ correspondingly (indicated by the solid line). 
 We can see that all the curves oscillate with $T$ due to the finite size effect. 
What is more, the oscillation period is almost on the same order for $E_1 = 3.368\times 10^{-5} meV$ at $B_x=2.25T$ and for $E_1 = 4.885\times 10^{-4} meV$ at $B_x=2.75T$.
 This is because $t_1$ also plays important role during the braiding. The values of the effective $t_1$ are fitted through the effective model of Eq. (\ref{Eq2}). As indicated by the corresponding dashed line in Fig. \ref{fig2}(b)-(d), the final braiding results are well captured by the effective model. Interestingly,  $t_1$ is closely related to  $E_1$. 
 As shown by the orange line in the bottom panel of Fig. \ref{fig2}(a), the effective  $t_1$
 also oscillates around zero energy but with a different period. Since the total oscillation period is determined by the interplay of the effects induced by $E_1$ and $t_1$,  the oscillation period of  $|\langle\psi_1^{-} (6T)|\psi_1^{+}(0)\rangle|$ is almost in the same order.

This conclusion can be further confirmed through spectroscopic analysis. 
Ref. [\onlinecite{AguadoN}] suggests that quantum dots at the ends of the nanowire can serve as a powerful tool to quantify the non-locality of MZMs. 
Our proposal involves a structure with a coupled quantum dot-nanowire system.
In such a structure, if $E_1\gg t_1$, the spectrum with respect to $E_d$  will display a  bowtie like shape, while if $t_1\gg E_1$, then the spectrum with respect to $E_d$ will display a diamond like shape. 
Figure 2(e)-(g) show the corresponding energy spectrum  versus the on-site energy of QD state $E_d$ with $C_{1}(t) = t_c$ and $C_{2}(t) = 0$.
It is indeed displays a bowtie like shape in Fig. \ref{fig2}(e) and Fig. \ref{fig2}(g). Here $E_1$ is almost zero while $t_1$ is very large. While the spectrum displays a diamond like shape in Fig. \ref{fig2}(f). 
Here $t_1$ is almost zero while $E_1$ is quite large. The spectrum is in fully consistent with the braiding results.

Since $t_1$ is closely related to  $E_1$,  they both oscillate around zero energy with the same order of magnitude but with a different periods.   
A plausible reason is that the MZMs are non-locally distributed at the two ends of the nanowire. The MZMs that are far from the QD connect to the QD through a finite length of the nanowire. 
We can expect that $ t_1$ is also determined by the length of the nanowire, which is similar to $E_1$. Thus, it is reasonable that \( E_1 \) and \( t_1 \) are of the same order. 
Following the 
Refs. [\onlinecite{Aguado2}],  we can quantify the non-locality of two MZMs through the parameters which named local estimator  $\eta = \sqrt{\frac{\left|u^2(x=L)\right|}{\left|u^1(x=L)\right|}}$.  Here $u^{1(2)}(x=L)$ is the normlized wavefunction of $\gamma_{1(2)}$ at the end.
The dark red line in the middle panel of Fig. \ref{fig2}(a) shows the corresponding curve of $\eta$  versus magnetic field $B_x$.  $\eta$ becomes very small when the system enters into the topological region. This suggests that MZMs is indeed non-locally distributed
at the two ends the wire.
Therefore, for a long nanowire, \( E_1 \) is exponentially small in the topological region, and the corresponding \( t_1 \) will also be very small.
 In such a situation, non-Abelian braiding would be sustained for a longer braiding time cost.

\begin{figure*}
\includegraphics[width=7in]{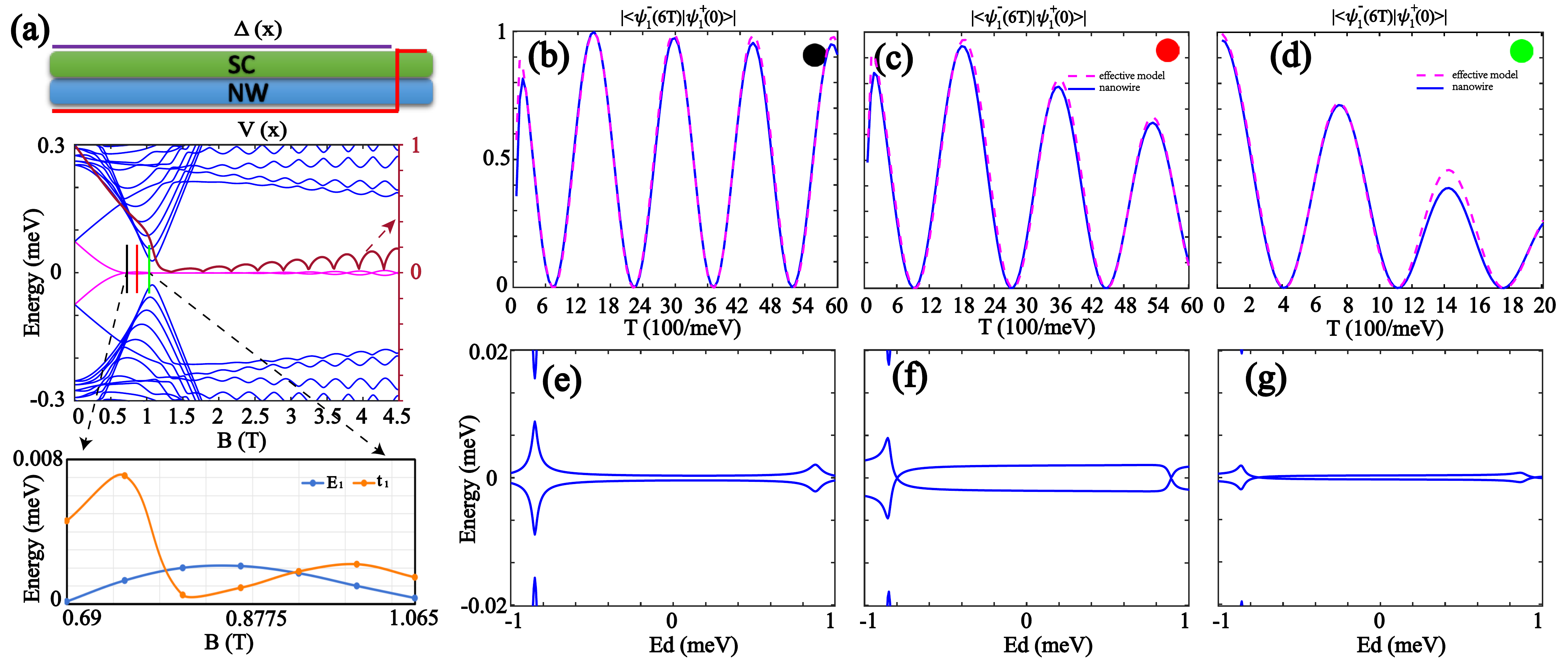}
\caption{\label{fig5} 
Step-like quantum-dot structure. (a) The schematic plot of the  step-like quantum-dot structure model is shown in the top panel. The middle panel shows the energy spectrum with $a = 20 nm$, and the dark red line shows the local estimator $\eta$. The bottom panel shows that the value of $E_1$ and $t_1$ as the functions of $B_x$. 
(b)-(d) show the braiding results as a function of braiding time cost $T$ at the different $B_x$，which corresponds to the values of $B_x$ marked by the vertical lines of different colors in the energy spectrum of (a)(indicated by the solid line) and the dashed line presents the fitting result by effective model.  
 (e)-(g) show the energy spectrum of QD-nanowire model as the function of the on-site energy of QD state $E_d$ at the different $B_x$, which corresponds to the values of $B_x$ marked by the vertical lines of different colors in the energy spectrum of (a).
 In the (b) and (e), $B_x = 0.6900 T$, $E_1 = 1.3569\times 10^{-4} meV$ and $t_1 = 4.60\times 10^{-3}meV$. 
 In the (c) and (f), $B_x = 0.8775 T$, $E_1 = 2.1000\times 10^{-3} meV$ and $t_1 = 9.00\times 10^{-4}meV$.
 In the (d) and (g), $B_x = 1.0650 T$, $E_1 = 3.2964\times 10^{-4} meV$ and $t_1 = 1.51\times 10^{-3}meV$.
 The other parameters are shown in the fourth row of Table \ref{Tab1}.
 }
\end{figure*}

\begin{figure*}
\includegraphics[width=7in]{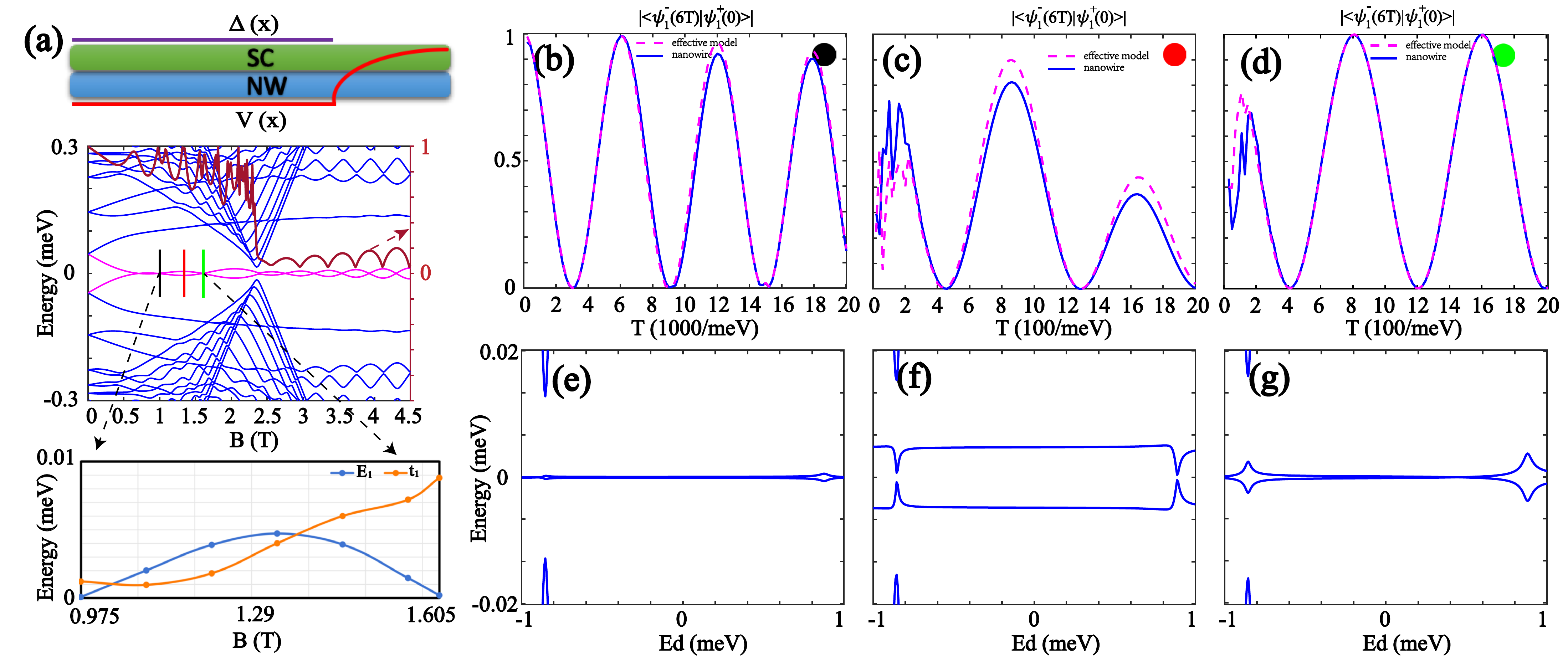}
\caption{\label{fig6}
Smooth quantum-dot structure. (a) The schematic plot of the smooth quantum-dot structure model is shown in the top panel. The middle panel shows the energy spectrum, and the dark red line shows the local estimator $\eta$. 
The bottom panel shows that the value of $E_1$ and $t_1$ as the functions of $B_x$. 
(b)-(d) show the braiding results as a function of braiding time cost $T$ at the different $B_x$，which corresponds to the values of $B_x$ marked by the vertical lines of different colors in the energy spectrum of (a)(indicated by the solid line) and the dashed line presents the fitting result by effective model.  
 (e)-(g) show the energy spectrum of QD-nanowire model as the function of the on-site energy of QD state $E_d$ at the different $B_x$, which corresponds to the values of $B_x$ marked by the vertical lines of different colors in the energy spectrum of (a).
 In the (b) and (e), $B_x = 0.975 meV$, $E_1 = 6.216\times 10^{-5} meV$ and $t_1 = 1.20\times 10^{-3}meV$. 
 In the (c) and (f), $B_x = 1.320 meV$, $E_1 = 4.717\times 10^{-3} meV$ and $t_1 = 4.00\times 10^{-3}meV$.
 In the (d) and (g), $B_x = 1.605 meV$, $E_1 = 1.896\times 10^{-4} meV$ and $t_1 = 8.80\times 10^{-3}meV$.
 The other parameters are shown in the fifth row of Table \ref{Tab1}.
 }
\end{figure*}

\subsection{\label{sec:B} Inhomogeneous potential at the boundary}
We have revealed the braiding properties in a finite-size semiconductor-superconductor system. We now aim to further study the braiding properties of ABSs under various different conditions. 
First, we consider the case where the chemical potential is inhomogeneous at the boundary.
As shown in the top panel of Fig. \ref{fig3}(a), the inhomogeneous chemical potential \( V(x) \) (indicated by the red line) exhibits a step-like potential at the ends, while the superconducting order parameter (indicated by the purple line) is uniform across the entire region. 
Since the ends or interfaces of a nanowire experience a different external field compared to the bulk, this case is quite common.
 As revealed by the energy spectrum in the middle panel of Fig.  \ref{fig3}(a),  the nearly zero-energy states emerge before the system entering into the topological region. 
   The corresponding local estimator, indicated by the dark red line in the middle panel of Fig. \ref{fig3}(a), is very large before the system entering into the topological region.  This suggests that
   the nearly zero energy states are confined within the inhomogeneous region and are not far from each other.
   Significantly, the braiding properties are not related with the local estimator.
 If we focus on the range from $1.4T$ to $1.9T$, these nearly zero energy states oscillate around zero energy as indicated by the blue line shown in the bottom panel of Fig. \ref{fig3}(a). 
 While  the orange solid line show the corresponding effective value of $t_1$,
which behaves similarly to the case with a finite length, regardless of the local estimator. 
 Fig.  \ref{fig3}(b)-(d) show the final weight of  $|\langle\psi_1^{-} (6T)|\psi_1^{+}(0)\rangle|$ versus the  time costing $T$ with $B_x=1.4T$, $B_x=1.6T$ and $B_x=1.9T$, correspondingly. The final braiding result oscillate almost on the same order. 
  This suggests that the braiding results are primarily related to $E_1$ and $t_1$.  The corresponding energy spectrum  versus the on-site energy of QD state $E_d$ at $C_{1}(t) = t_c$ and $C_{2}(t) = 0$ in Fig.  \ref{fig3} (e)-(g) is also consistent
 with the braiding results. 
 
 We further consider the case that the inhomogeneous potential $V(x)$ has a smooth spatial profile, as described by the parameters in the third row of Table  \ref{Tab1}. 
 In this situation, the energy spectrum of the ABSs is closer to zero energy and about one-tenth of that in the case where  $V(x)$ is a step potential.
Interestingly, the corresponding  effective $t_1$ is of the same order of $E_1$, as indicated by the orange line in the bottom panel of Fig.  \ref{fig4}(a). 
 The oscillation period of  $|\langle\psi_1^{-} (6T)|\psi_1^{+}(0)\rangle|$, shown in Fig. \ref{fig4}(b)-(d), also increases and becomes about ten times greater than that in the case where  $V(x)$ is a step potential.
However, the local estimator indicated by the dark red line in the middle bottom panel of Fig. \ref{fig4}(a) remains the same as before.
By comparing the braiding properties at two different energy scales, $E_1$, while maintaining the same locality, we can observe that the braiding property is mainly associated with the stability of $E_1$.
In other words, if the fluctuation amplitude of $E_1$ around zero energy  is small, then the fluctuation amplitude of $t_1$ will also be small. Then the non-abelian braiding will sustain for a longer braiding time cost.
The braiding property of these nearly zero-energy states behaves similarly to that of two coupled Majorana zero modes (MZMs) in a finite length, but with a lower non-locality. 
A possible reason might be related to the spin degrees of freedom of MZMs in a nanowire.
Wimmer et.al.  have demonstrated that two Majorana zero modes (MZMs) with finite overlap have opposite spins. 
Consequently, these two MZMs will experience different tunnel barriers when traversing an inhomogeneous potential.
 This results in an exponential difference in the coupling between the two MZMs and the environment \cite{Wimmer}. 
 Our braiding results are fully consistent with this theory.

\begin{figure*}
\includegraphics[width=7in]{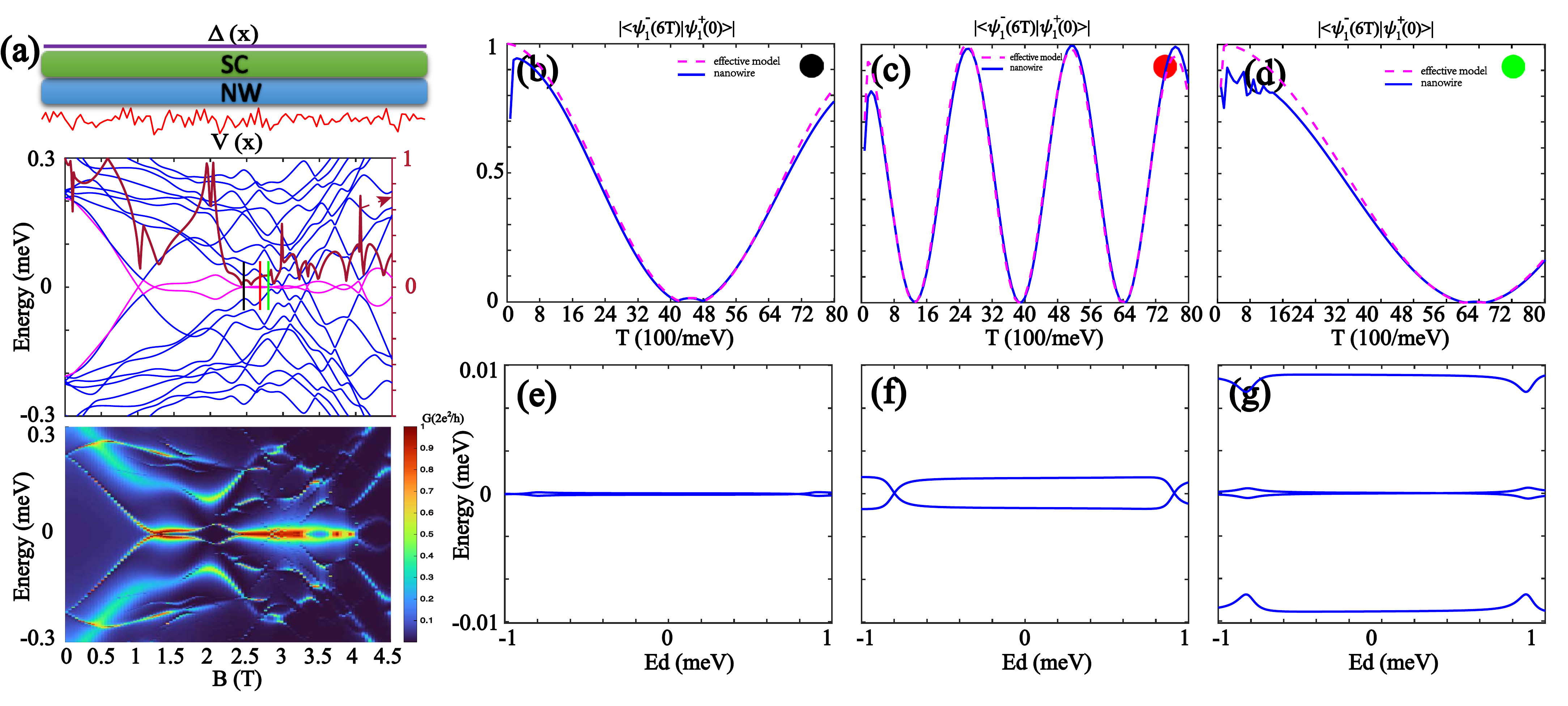}
\caption{\label{fig7}
Disorder. (a) The schematic plot of the disorder model is shown in the top panel. The middle panel shows the energy spectrum, and the dark red line shows the local estimator $\eta$. The bottom panel shows the conductance with $\boldsymbol{\Sigma} = 0.3 meV$. 
(b)-(d) show the braiding results as a function of braiding time cost $T$ at the different $B_x$，which corresponds to the values of $B_x$ marked by the vertical lines of different colors in the energy spectrum of (a)(indicated by the solid line) and the dashed line presents the fitting result by effective model.  
 (e)-(g) show the energy spectrum of QD-nanowire model as the function of the on-site energy of QD state $E_d$ at the different $B_x$, which corresponds to the values of $B_x$ marked by the vertical lines of different colors in the energy spectrum of (a).
 In the (b) and (e), $B_x = 2.484 T$, $E_1 = 5.593\times 10^{-5} meV$ and $t_1 = 4\times 10^{-4}meV$. 
 In the (c) and (f), $B_x = 2.715 T$, $E_1 = 1.200\times 10^{-3} meV$ and $t_1 = 2\times 10^{-4}meV$.
 In the (d) and (g), $B_x = 2.807 T$, $E_1 = 2.247\times 10^{-5} meV$ and $t_1 = 5\times 10^{-4}meV$.
 The other parameters are shown in the sixth row of Table \ref{Tab1}.
}
\end{figure*}

%
%
%
%
%

\subsection{\label{sec:C} Quantum-dot like structure at the boundary}
Since superconductivity is induced through the proximity effect in a nanowire-superconductor hybrid system, it is likely that superconductivity is weakened at the interface.
In this scenario, both the chemical potential and the superconducting amplitude are inhomogeneous at the interface. 
This creates an additional quantum-dot-like structure at the boundary. Consequently, sub-gap states may form due to this quantum-dot structure .
Such sub-gap states could evolve into nearly zero-energy states with the variations of the magnetic field under certain conditions.
 In the middle panel of Fig. \ref{fig5}(a),  the energy spectrum displays a nearly zero-energy state before the system entering into
the topological region with the parameters shown in the fourth row of the Table  \ref{Tab1}.  Fig.  \ref{fig5}(b)-(d) show the corresponding braiding results  $|\langle\psi_1^{-} (6T)|\psi_1^{+}(0)\rangle|$ versus braiding time cost $T$. 
Clearly, the braiding results oscillate with \( T \) and are well captured by the effective model. We further quantify the corresponding \( t_1 \) using the effective model, with the data indicated by the orange line in the bottom panel of Fig. \ref{fig5}(a). The behavior of the nearly zero-energy states remains consistent with previous observations. The corresponding \( E_1 \) and \( t_1 \) fluctuate around zero energy and effectively compensate with each other. 
As a result, the oscillation period of  $|\langle\psi_1^{-} (6T)|\psi_1^{+}(0)\rangle|$ is nearly the same order within a finite range and determined by the stability of $E_1$.

We also consider a quantum dot structure where the chemical potential is spatially smooth. The corresponding parameters are provided in the fifth line of Table 1. As shown in Fig.  \ref{fig6}, the oscillation period of \( |\langle \psi_1^{-} (6T) | \psi_1^{+}(0) \rangle | \) is still determined by the stability of \( E_1 \) . There is no significant difference whether the chemical potential spatially varies rapidly or smoothly. 
This suggests that the nearly zero-energy states in a quantum-dot-like structure can also be attributed to finite coupled MZMs.
The Andreev bound states (ABSs) in a quantum-dot-like structure might also be utilized for topological quantum computation if the fluctuation of \( E_1 \) is small. However, stabilizing ABSs in such a structure is more challenging compared to the cases where only the chemical potential is inhomogeneous.

\subsection{\label{sec:D}Disorder at the bulk}
In the state art of nanotechnology, disorder inevitably occurs in nanowire systems. 
Since the first semiconductor-superconductor nanowires were fabricated experimentally in 2012, this issue has attracted significant attention.
Theoretical and experimental studies suggest that disorder can lead to the collapse of bulk states and induce similar zero-bias peaks (ZBPs) in experiments.
Although disorder has been largely supressed and quantized-ZBPs have been observed experimentally over the past decade, theory suggests that quantized-ZBPs may still be detected in trivial situations under the influence of disorder. Here we investigate the braiding properties of MZMs in the presence of disorder. We focus mainly on the ABSs which can  cause  quantized-ZBPs in the conductance measurement.
 In the middle panel of Fig.  \ref{fig7}(a), a typical energy spectrum resulting from disorder is shown. In this case, $V(x)$ is a random potential represented by an uncorrelated Gaussian distribution, i.e., $V(x) \sim \mathcal{N}\left(0,1^2\right)$, where $V(x) \sim \mathcal{N}\left(0,1^2\right)$ denotes a Gaussian distribution with mean value of 0 and variance of 1.
 The bottom panel of Fig.  \ref{fig7}(a) displays the corresponding differential conductance as a function of magnetic field and voltage. We observe that a quantized-ZBPs appears within a specific range.
The range of the quantized-ZBPs is not fully consistent with the spectrum because some of the nearly zero-energy ABSs are confined within the bulk region due to disorder. Here, we focus on the range where ZBPs can be detected. Figure \ref{fig7} (b)-(d) show  $|\langle\psi_1^{-} (6T)|\psi_1^{+}(0)\rangle|$ versus the time
cost $T$ at some typical position. Interestingly, we find that these ABSs exhibit behavior similar to that of finite-overlap MZMs. Their properties are influenced by the interplay between the effect induced by $E_1$ and $t_1$. 
Moreover, $t_1$ is also related with $E_1$ and they both oscillate with the oscillation period almost on the same order. This further suggests that the stability of non-abelian braiding is mainly determined by the stability of $E_1$.  To show the generality of this conclusion, we calculate another configuration of disorder, the results are presented in Appendix C. This suggests that these nearly zero-energy states may also originate from the inhomogeneous potential at the interface. Disorder could induce a local potential profile that behaves similarly to the inhomogeneous potential at the interface. Consequently, the ABSs induced by disorder behave similarly to those found in the presence of an inhomogeneous potential.

\begin{figure*}
\includegraphics[width=7in]{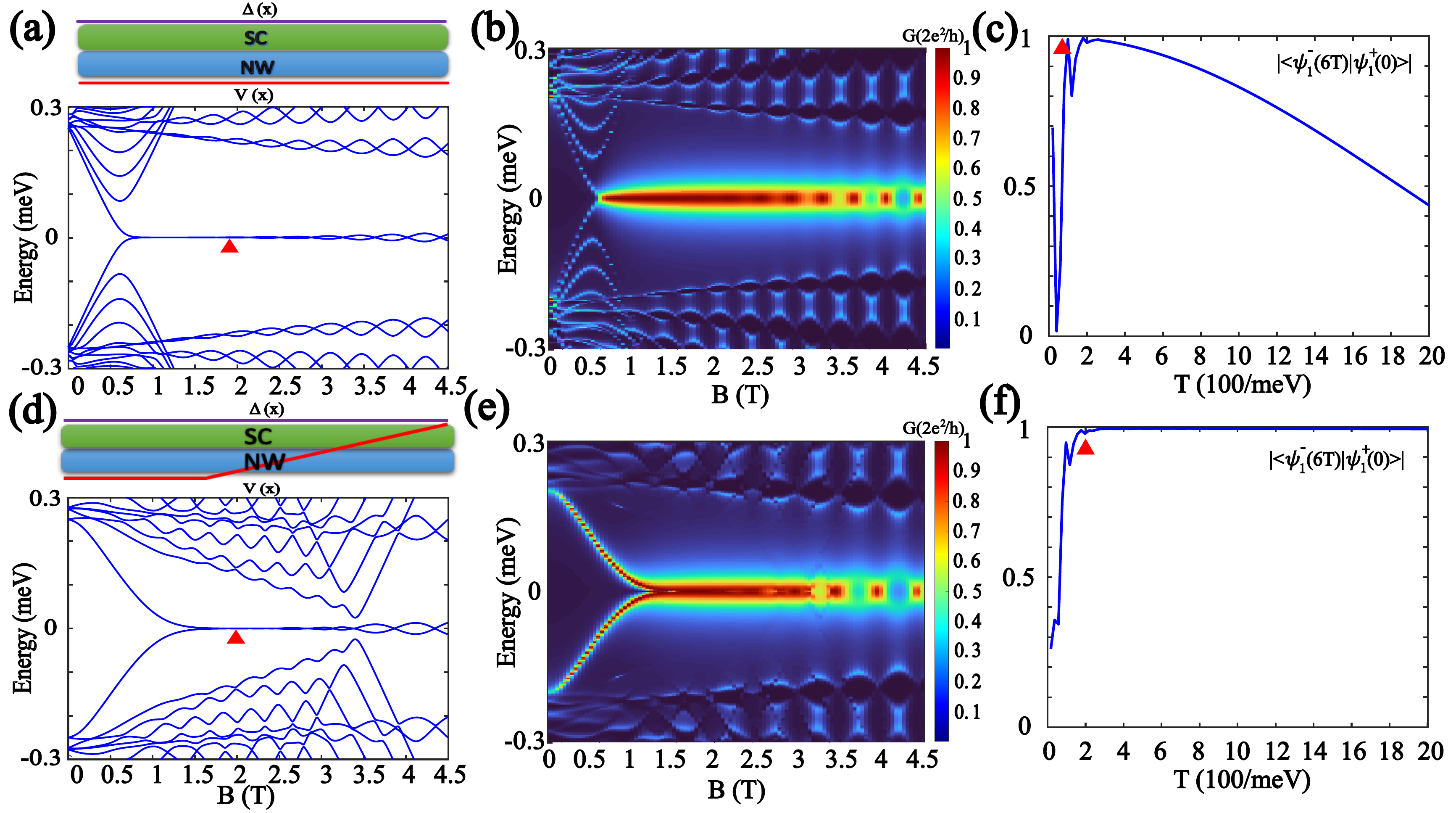}
\caption{\label{fig8}
(a) The schematic plot of the uniform nanowire with $L=2.0 \mu m$ is shown in the top panel. The bottom panel shows the energy spectrum. (b) The schematic plot of the inhomogeneous potential with $L=2.0 \mu m$ is shown in the top panel. The bottom panel shows the energy spectrum. (b) and (e) show the conductance corresponding to (a) and (d), respectively. And $\boldsymbol{\Sigma} = 0.3 meV$. (c) and (e) show the braiding results as a function of braiding time cost at the magnetic filed $B_x = 1.9T$ indicated by the red triangle in (a) and (d).
 }
\end{figure*}

\subsection{\label{sec:E}Non-Abelian braiding modified by  inhomogeneous potential}

We have revealed that these nearly zero energy ABSs, despite differing mechanisms, exhibit properties similar to those of finite-overlap MZMs. 
Their non-abelian braiding property will be gradually destroyed by additional $E_1$ and $t_1$. Since $t_1$ is also related to $E_1$, the primary criterion is actually  the stability of $E_1$ near zero energy. 
If $E_1$ fluctuates with a vanishingly small amplitude around zero energy, the oscillation period of the braiding results will be quite large. In this situation, if the braiding process is completed with $T\ll 1/E_1$, 
the dynamical effects can be neglected, leading to high fidelity in the braiding.
This suggests  that these nearly zero energy ABSs might still be  used for non-Abelian braiding if we can stabilize $E_1$ at zero energy. Experimentally, the typical length of a nanowire is about $2\mu m$. 
Fig. \ref{fig8}(a) shows the corresponding energy spectrum of a nanowire of this length as a function of the magnetic field.
 In such a clean nanowire system, it will enter into topological region if $B>1T$. The differential conductance in Fig. \ref{fig8}(b) exhibits a clear QZBP in the topological region.
 We further investigate the non-Abelian braiding properties of MZMs at $B=1.9T$. 
Although swapping is successful if the time cost $T$ is short, Fig. \ref{fig8}(c) reveals that deviations from perfect swapping occur as $T$ increases, eventually leading to oscillations with $T$. 
Interestingly, if we introduce an inhomogeneous potential at the interface while keeping other parameters unchanged, quasi-Majorana zero modes (quasi-MZMs) will emerge, as shown in Fig. \ref{fig8}(d). 
These quasi-MZMs will also induce stable QZBPs in the differential conductance \cite{155314, 035312, 184520, 013377}.
Moreover, the non-Abelian braiding fidelity is significantly enhanced in this situation. Fig. \ref{fig8}(f) demonstrates that perfect swapping can be sustained for longer braiding time cost without any deviation. 
This suggests that these nearly zero energy ABSs may also be suitable for topological quantum computation.

\section{\label{sec:5}Conclusion}
We have systemically  investigated the braiding properties of ABSs induced by various inhomogeneous potential in the nanowire. 
We find that these ABSs can be considered as MZMs with finite overlap.
These finite-overlap Majorana zero modes introduce additional coupling terms $E_1$ and $t_1$, which lead to dynamical effects and deviations from the non-Abelian braiding properties of MZMs.
The numerical simulations are in good agreement with the effective model.
Moreover, the deviation is influenced by the stability of $E_1$ within a certain range. In other words, if $E_1$ remains at zero energy with vanishingly small fluctuation, the non-Abelian braiding properties will maintain over an extended range.
Indeed, We found  the non-abelian braiding of ABS under certain conditions can outperform that of MZMs in realistic systems. 
This suggests that ABSs may also be suitable for topological quantum computation.  

Finally, we want to emphasize that quantized zero-bias peaks could still be a valuable tool for distinguishing Majorana zero modes (MZMs). The braiding properties of Andreev Bound States (ABSs) suggest that these nearly zero-energy states may be attributed to finite-overlap Majorana zero modes. 
The non-Abelian braiding property of these nearly zero-energy states can remain stable for a longer braiding time cost if the fluctuation of \( E_1 \) around zero energy is tiny. This implies that if quantized zero-bias peaks are observed across a certain range, then this range could be a promising platform for non-Abelian braiding. To further quantify the stability of these ABSs, additional operations such as performing the braiding and fusion protocols may be required.

\begin{acknowledgments}
This work is financially supported by National Natural Science Foundation of China (Grants No. 92265103 and No. 12304194), and the Innovation Program for Quantum Science and Technology (Grant No. 2021ZD0302400).
\end{acknowledgments}

\newpage
\appendix

\renewcommand{\theequation}{S\arabic{equation}}
\renewcommand{\thefigure}{S\arabic{figure}}
\renewcommand{\bibnumfmt}[1]{[S#1]}
\renewcommand{\citenumfont}[1]{S#1}

\section{Local estimator $\eta$}

Consider a general subgap eigenstate in a nanowire system:
\begin{equation}
d=\int d x \sum_\sigma u_\sigma(x) \psi_\sigma(x)+v_\sigma(x) \psi_\sigma^{\dagger}(x)
\end{equation}

Here, $u_\sigma(x)$ and $v_\sigma(x)$ is the probability amplitude of the particle and hole wave function. $\sigma$ is the spin projections. We can decompose this fermion state into two Majorana components $\gamma_L$ and $\gamma_R$.
\begin{equation}
\begin{aligned}
d & =\frac{\gamma_L+i \gamma_R}{\sqrt{2}} \\
d^{\dagger} & =\frac{\gamma_L-i \gamma_R}{\sqrt{2}}
\end{aligned}
\end{equation}

so that,
$\gamma_{1, 2}=\int d x \sum_\sigma u_\sigma^{1,2}(x) \psi_\sigma(x)+\left[u_\sigma^{1, 2}(x)\right]^* \psi_\sigma^{\dagger}(x)$.
The $\gamma_{1,2}$ is self-conjugate. The probability amplitude of the $\gamma_{1,2}$ can be written as:
\begin{equation}
\begin{aligned}
& u_\sigma^1(x)=\frac{u_\sigma(x)+v_\sigma^*(x)}{\sqrt{2}} \\
& u_\sigma^2(x)=\frac{u_\sigma(x)-v_\sigma^*(x)}{i \sqrt{2}}
\end{aligned}
\end{equation}

Thus, we can define the local estimator $\eta$:
\begin{equation}
\eta = \sqrt{\frac{\left|u^1(x=L)\right|}{\left|u^2(x=L)\right|}}
\end{equation}

\section{Differential conductance}
To simulate the experimental measurement of tunneling conductance G, we attach a normal lead to the end of the nanowire and numerically calculate the tunneling conductance by Non-equilibrium Green's function.
\begin{equation}
\mathbf{G}^r(\epsilon)=1 /\left[\epsilon \mathbf{I}-\mathbf{H}_{s}-\sum_m \boldsymbol{\Sigma}_m^r\right]
\end{equation}

Here, the linewidth function can be computed through:
\begin{equation}
\boldsymbol{\Gamma}_m(\epsilon)=i\left[\boldsymbol{\Sigma}_m^r(\epsilon)-\boldsymbol{\Sigma}_m^a(\epsilon)\right]
\end{equation}

In the subsequent calculation, we set the self-energy to a constant.

\section{Disorder at the bulk}
\begin{figure*}
\includegraphics[width=7in]{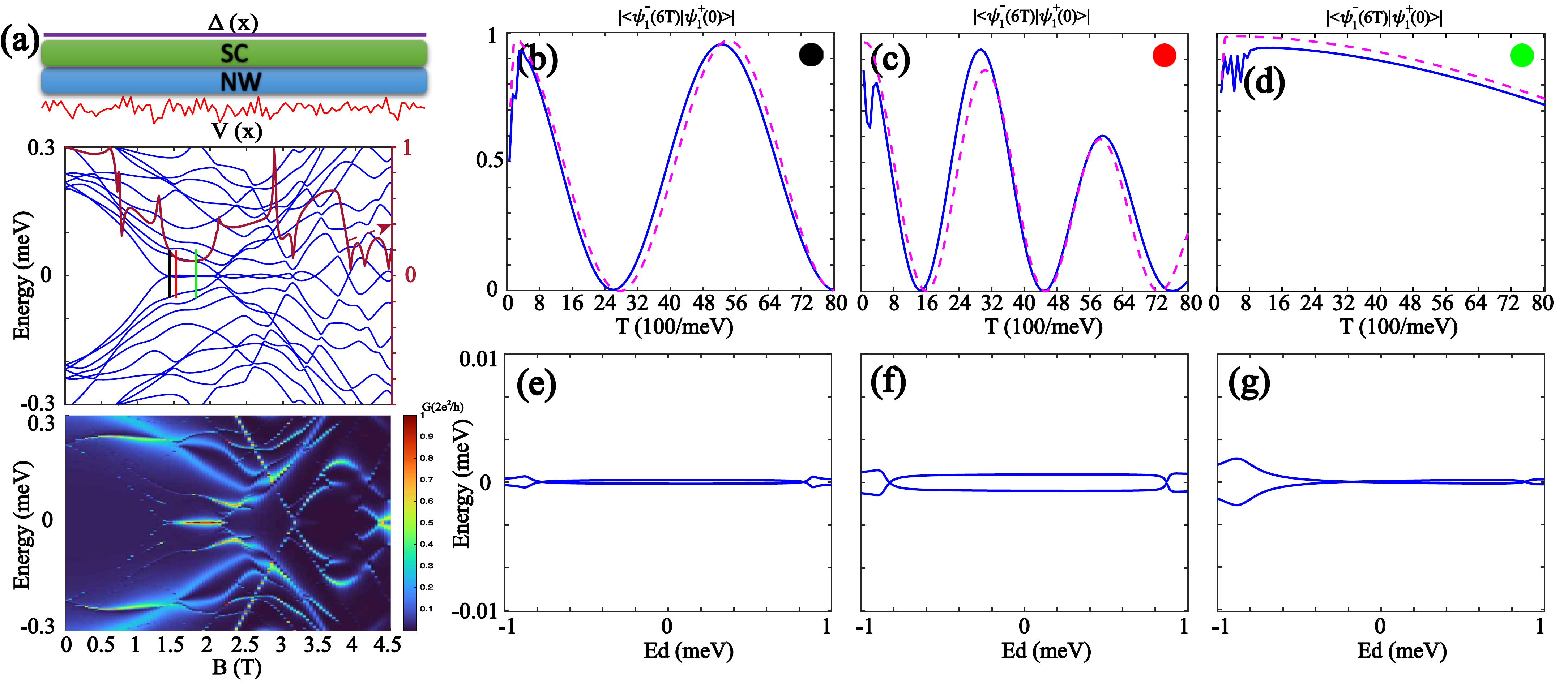}
\caption{\label{fig9}Disorder. (a) The schematic of the disorder model is shown in the top panel. The middle panel shows the energy spectrum, and the dark red shows the local estimator $\eta$. The bottom panel shows the conductance with $\boldsymbol{\Sigma} = 0.3 meV$. 
(b)-(d) show the braiding results as a function of braiding cost time at the different $B_x$，which corresponds to the values of $B_x$ marked by the vertical lines of different colors in the energy spectrum of (a)(indicated by the solid line) and the dashed line presents the fitting result by effective model.  
 (e)-(g) show the energy spectrum of QD-nanowire model as the function of the on-site of QD state $E_d$ at the different $B_x$, which corresponds to the values of $B_x$ marked by the vertical lines of different colors in the energy spectrum of (a).
 In the (b) and (e), $B_x = 1.84 T$, $E_1 = 7.88\times 10^{-5} meV$ . 
 In the (c) and (f), $B_x = 1.93 T$, $E_1 = 6.71\times 10^{-4} meV$ .
 In the (d) and (g), $B_x = 2.02 T$, $E_1 = 5.75\times 10^{-5} meV$ .
 The other parameters are shown in the sixth row of Table \ref{Tab1}.
 }
\end{figure*}

we calculate another configuration of disorder, the results are the same as before. 
Fig. \ref{fig9} (b)-(d) show $|\langle\psi_1^{-} (6T)|\psi_1^{+}(0)\rangle|$ versus the braiding  time
cost $T$ at some typical position. These ABSs exhibit behavior similar to that of finite-overlap MZMs. Their properties are influenced by the interplay between the effects induced by $E_1$ and $t_1$. 
 $t_1$ is also related with $E_1$ and they both oscillate with almost the same order of oscillation period.

\end{CJK} 
\end{document}